\documentclass[aps,prl,floatfix,superscriptaddress,nofootinbib,showpacs,showkeys,preprintnumbers]{revtex4}

\usepackage[dvips]{epsfig}
\usepackage{graphicx}

\begin{document}

\preprint{PREPRINT NPI MSU 2004-4/743}

\title{Observation of narrow baryon resonance decaying into $pK^0_s$ in $pA$-interactions at $70\ GeV/c$ with SVD-2 setup.}

\def\groupsinp{\affiliation{D.V. Skobeltsyn Institute of Nuclear Physics, Lomonosov Moscow State University, 1/2 Leninskie gory, Moscow, 119992 Russia}}
\def\groupihep{\affiliation{Institute for High-Energy Physics, Protvino, Moscow oblast,  142284, Russia}}
\def\groupjinr{\affiliation{Joint Institute for Nuclear Research, Dubna, Moscow oblast, 141980, Russia}}
\def\groupzel{\affiliation{Research Institute of Materials Science and Technology, 103460, Moscow, Zelenograd, Russia}}
\def\groupzell{\affiliation{NPO "NIITAL",  103460, Moscow, Zelenograd, Russia}}

\groupsinp
\groupihep
\groupjinr
\groupzel
\groupzell

\author{A.~Aleev} \groupjinr
\author{N.~Amaglobeli} \groupjinr
\author{E.~Ardashev} \groupihep
\author{V. ~Balandin} \groupjinr
\author{S.~Basiladze} \groupsinp
\author{S.~Berezhnev} \groupsinp
\author{G.~Bogdanova} \groupsinp
\author{M.~Boguslavsky} \groupjinr
\author{N.~Egorov} \groupzel
\author{V.~Ejov} \groupsinp
\author{G.~Ermakov} \groupsinp
\author{P.~Ermolov} \groupsinp
\author{N.~Furmanec} \groupjinr
\author{S.~Golovnia} \groupihep
\author{S.~Golubkov} \groupzel
\author{A.~Gorkov} \groupzell
\author{S.~Gorokhov} \groupihep
\author{I.~Gramenitsky} \groupjinr
\author{N.~Grishin} \groupsinp
\author{Ya.~Grishkevich} \groupsinp
\author{D.~Karmanov} \groupsinp
\author{A. ~Kholodenko} \groupihep
\author{A.~Kiriakov} \groupihep
\author{I.~Kosarev} \groupjinr
\author{N.~Kouzmine} \groupjinr
\author{V.~Kozlov} \groupsinp
\author{Yu.~Kozlov} \groupzel
\author{E.~Kokoulina} \groupjinr
\author{N.V.~Korotkov} \groupsinp
\author{V.~Kramarenko} \groupsinp
\author{A.~Kubarovsky\footnote[1]{Contact person, e-mail: alex\_k@hep.sinp.msu.ru}} \groupsinp
\author{L.~Kurchaninov} \groupihep
\author{V.~Kuzmin} \groupsinp
\author{E.~Kuznetsov} \groupsinp
\author{G.~Lanshikov} \groupjinr
\author{A.~Larichev} \groupsinp
\author{A.~Leflat} \groupsinp
\author{M.~Levitsky} \groupihep
\author{S.~Lyutov} \groupsinp
\author{S.~Maiorov} \groupsinp
\author{M.~Merkin} \groupsinp
\author{A.~Minaenko} \groupihep
\author{G.~Mitrofanov} \groupihep
\author{A.~Moiseev} \groupihep
\author{V.~Murzin} \groupsinp
\author{V.~Nikitin} \groupjinr
\author{P.~Nomokonov} \groupsinp
\author{A.~Oleinik} \groupjinr
\author{S.~Orfanitsky} \groupsinp
\author{V.~Parakhin} \groupihep
\author{V.~Petrov} \groupihep
\author{L.~Pilavova} \groupzell
\author{A.~Pleskach} \groupihep
\author{V.~Popov} \groupsinp
\author{V.~Riadovikov} \groupihep
\author{R.~Rudenko} \groupihep
\author{I.~Rufanov} \groupjinr
\author{V.~Senko} \groupihep
\author{M.~Shafranov} \groupjinr
\author{N.~Shalanda} \groupihep
\author{A.~Sidorov} \groupzel
\author{M.~Soldatov} \groupihep
\author{L.~Tikhonova} \groupsinp
\author{T.~Topuria} \groupjinr
\author{Yu.~Tsyupa} \groupihep
\author{M.~Vasiliev} \groupihep
\author{A.~Vischnevskaya} \groupsinp
\author{V.~Volkov} \groupsinp
\author{A.~Vorobiev} \groupihep
\author{A.~Voronin} \groupsinp
\author{V.~Yakimchuk} \groupihep
\author{A.~Yukaev} \groupjinr
\author{L.~Zakamsky} \groupihep
\author{V.~Zapolsky} \groupihep
\author{N.~Zhidkov} \groupjinr
\author{V. ~Zmushko} \groupihep
\author{S.~Zotkin} \groupsinp
\author{D.~Zotkin} \groupsinp
\author{E.~Zverev} \groupsinp

\collaboration{The SVD Collaboration} \noaffiliation

\date{\today. ~To be submitted to Yadernaya Fizika}

\begin{abstract}
SVD-2 experiment data have been analyzed to search for an exotic baryon state,
the $\Theta^+$-baryon, in a $pK^0_s$ decay mode at $70~GeV/c$ on IHEP accelerator. The reaction
$pA \to pK^0_s+X$ with a limited multiplicity was used in the analysis.
The $pK^0_s$ invariant mass spectrum shows a resonant structure with $M=1526\pm3(stat.)\pm 3(syst.)~MeV/c^2$ and $\Gamma < 24~MeV/c^2$. The statistical significance
of this peak was estimated to be of $5.6~\sigma$. The mass and width
of the resonance is compatible with the recently reported $\Theta^+$- baryon
with positive strangeness which was predicted as an exotic pentaquark ($uudd\bar{s}$) baryon state. The total cross section for $\Theta^+$ production in pN-interactions for $X_F\ge 0$ was estimated to be $(30\div120)~\mu b$ and no essential deviation from A-dependence for inelastic events $(\sim A^{0.7})$ was found.
\end{abstract}

\maketitle

\section {Introduction.}

Exotic baryons with 5-quarks content (pentaquarks) and their properties have been predicted by Diakonov, Petrov, and Polyakov \cite{Diakonov} in the framework of the chiral
soliton model, although
such 5-quarks structures were proposed years ago \cite{jaffe1,pras}.
The lightest member of the pentaquarks antidecuplet, $\Theta^+$-baryon  predicted in \cite{Diakonov}, has positive strangeness, mass  $M \sim 1530~MeV/c^2$,~$\Gamma \le 15~MeV/c^2$, spin $1/2$ and even parity. Later, stable $uudd\bar{s}$ pentaquarks in the constituent quark model were suggested by Stancu and Riska \cite{stancu}.  Capstick {\it et al.} \cite{capstick} proposed the interpretation of the $\Theta^+$ as an isotensor pentaquark, Karliner and Lipkin have developed a cluster model using a diquark-triquark configuration\cite{lipkin}, in which the $\Theta^+$ is a positive-parity isosinglet member of antidecuplet.
Jaffe and Wilczek have suggested an underlying quark model structure
of this state \cite{jaffe}. Also there are attempts to predict $\Theta^+$-baryon with negative parity using lattice QCD \cite{qcd1,qcd2}.

Experimental evidence for $\Theta^+$-baryon with positive strangeness
came recently from several experimental groups (LEPS\cite{Nakano}, DIANA-ITEP \cite{Dolgolenko,itep}, CLAS \cite{Ken,vpk,clas2}, SAPHIR \cite{SAPHIR}). In those experiments $\Theta^+$-baryon was observed in the $nK^+$ or $pK^0_s$ invariant mass spectra with the mass near $1540~MeV/c^2$. More recently HERMES collaboration observed narrow baryon state at the mass of $1528\ MeV/c^2$ in quasi-real photoproduction\cite{hermes}. Also ZEUS collaboration \cite{zeus} reported an evidence of the exotic baryon in $pK^0_s$-channel with the mass of $1527\ MeV/c^2$.

In this paper we present the preliminary results of our searches for $\Theta^+$-baryon in proton-nuclear interactions (C,Si,Pb) at
$70\ GeV/c (\sqrt{S}=11.5\ GeV/c)$ on IHEP accelerator (Protvino) with the SVD-2 setup in the reaction: \\

$pN\rightarrow \Theta^+ + X$, \ $\Theta^+ \rightarrow pK^0_s$, \  $K^0_s \rightarrow \pi^+\pi^-$. \\

The data analysis was done in the inclusive reactions with the limited multiplicity in the beam proton fragmentation region ($X_F(pK^0_s)>0$).

\section {SVD-2 experimental setup.}

The main goal of SVD-2 experiment is a study of the charm hadroproduction at the near-threshold energy \cite{svd0,svd1,svd2,svd3}.

\begin{figure}[ht]
\vspace{80mm}
{\includegraphics{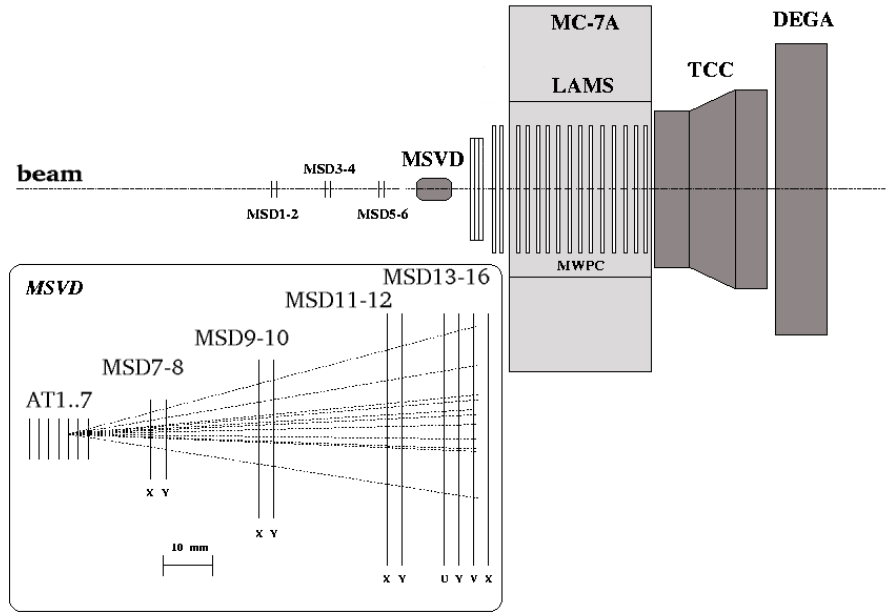}}
\caption{SVD-2 layout.}
\label{k0}
\end{figure}

The SVD-2 layout is presented on Fig. 1. The SVD-2 setup includes the following basic components:
\begin{enumerate}
\item The high-presicion microstrip vertex detector(MSVD). MSVD has the following elements:
\begin{itemize}
\item MSBT(microstrip beam telescope) - 3 XY pairs of microstrip Si-detectors (MSD1-6 each having 128 strips with 50$~\mu m$-pitch ), measuring XY-coordinates.
\item AT(active target) - 5 Si-detectors with thickness of $300~\mu m$,
Pb-foil with thickness of $220~\mu m$ and C-target with thickness of $500~\mu m$. The distance between all detectors is $4\ mm$.
\item MSVD(microstrip vertex detector) - 3 XY pairs of microstrip Si-detectors - MSD7-8 of 640 strips/$25~\mu m$-pitch, MSD9-10 (640/50 $~\mu m$), MSD11-12 (1024/50$~\mu m$) and UYVX quadruplet, MSD13-16 (1024/50$~\mu m$).
\end{itemize}
\item Large aperture magnetic spectrometer(LAMS) has the following main elements:
\begin{itemize}
\item Electromagnet MC-7A with aperture of $1.8\times1.3\ m^2$ and homogenous field of $1.18\ T$ over $3\ m$ long region.
\item A two sets of MWPC(multiwire proportional chambers). First one consists of 1 UYV triplet (sensitive area $1.0\times1.0\ m^2$ and $2~mm$ interwire distances) and placed before magnet, the second one (between magnet poles) consists of 5 UYV triplets with 2mm interwire distances and sensitive area of $1.0\times1.5\ m^2$. The part of these chambers are placed in the weak magnetic field.
\end{itemize}
\item The multicell threshold Cherenkov counter(TCC). The counter is designed for charged particles identification, it has an entrance aperture of $177\times 130\ cm^2$. Four rows with 8 spherical mirrors in each one are arranged on the rear wall of the counter. The threshold momentum for pions and protons is $4~GeV/c$ and $21~GeV/c$ respectivetly. When filled with freon at the atmospheric pressure and the temperature of $20 ^\circ C$, the counter provides the identification of pions in momentum range $4-21\ GeV/c$ with $70\%$ efficiency.
\item The gamma quanta detector (DEGA). DEGA consists of 1536 Cherenkov full absorption lead glass counters with the transverse dimensions of $38\times 38\ mm^2$ and the length of $505\ mm$. The total sensitive area of the detector is $1.8\times 1.2\ m^2$. DEGA provides a $\gamma$ registration with energies from $50\ MeV$ to $20\ GeV$ with the position resolution of $2-3\ mm$.
\end{enumerate}

\begin{figure}[ht]
\vspace{80mm}
{\includegraphics{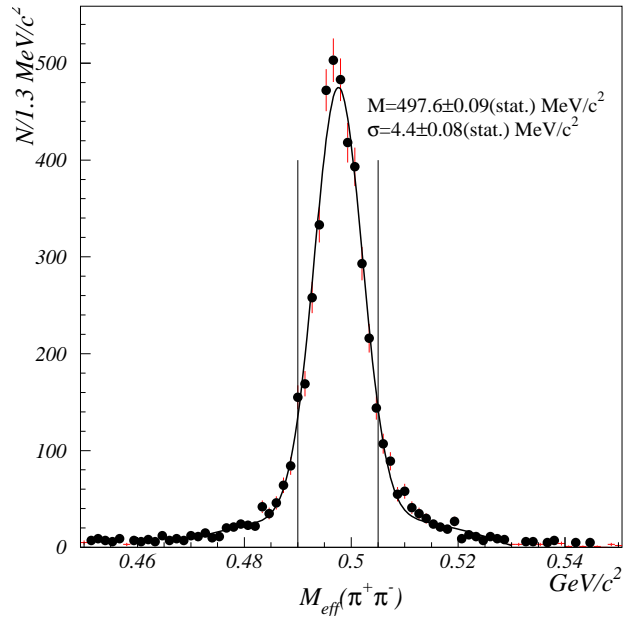}}
\caption{
The $(\pi^+\pi^-)$ invariant mass spectrum. A window corresponding to $\pm2\sigma$ is shown by the vertical lines.}
\label{k0}
\end{figure}

\begin{figure}[ht]
\vspace{80mm}
{\includegraphics{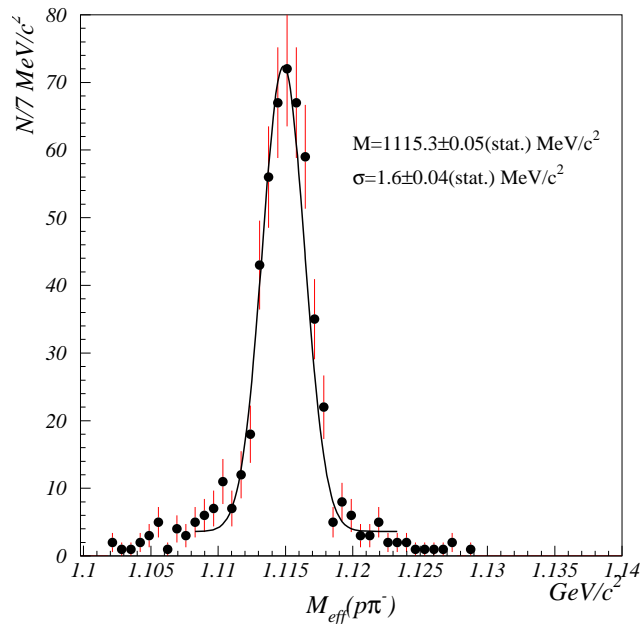}}
\caption{
The $(p \pi^-)$ invariant mass spectrum.}
\label{lambda}
\end{figure}

The SVD-2 trigger system provides the trigger signal, based on data from scintillation counters before active target, from scintillation hodoscope, situated after TCC and from the ionization losses, registered in the active target of MSVD. Every silicon plate of the active target is segmented by eight strips. The trigger electronics consists of three-level comparator modules accepting
analog signal from every strip of the active target and lookup-table module accepting digitizated data from the comparator output. Two-level lookup-table module can be turned to select the events with primary vertices in any of AT planes.

The SVD-2 spectrometer works in the proton beam of IHEP accelerator (U-70) with energy $E_p = 70\ GeV$ and intensity $I \approx (5 \div 6) \cdot 10^5$ p/cycle. The total statistics of $5\cdot10^7$ inelastic events was obtained. The integrated luminosity of this experiment for inelastic $pN$-interactions taking in account the triggering efficiency was $1600\ events/~\mu b$.

The primary vertex position determination procedure is based on well-known "tear-down" approach \cite{tear1, tear2}. First of all, the straight lines are drawn through the XY-counts in microstrip tracker by the least squares method. Several quality cuts are applied to the tracks reconstructed in an event, in order to remove fake or badly reconstructed tracks. Then
all survived tracks are included in the vertex fit and $\chi^2(N_{tracks})$ is computed.
Next, each track is excluded separately and a new $\chi^2(N_{tracks}-1)$ is computed. One then selects the track which gives maximum difference $\chi^2(N_{tracks})-\chi^2(N_{tracks}-1)$ and excludes it from the fit if the difference exceeds some threshold $\Delta_{max}$. This procedure is repeated while there remain the tracks to be excluded.
Only events with a good reconstructed primary vertex are selected.
After excluding the tracks that belongs to primary vertex, the secondary vertex position is determinated by finding V0-decay downstream the primary vertex on both X and Y-projections of the vertex tracker. The primary vertex resolution was estimated as $70-120~\mu m$ for Z-coordinate and $8-12~\mu m$ for X(Y)-coordinates. For the secondary vertices ($K^0_s,\ \Lambda$)  those values were $250~\mu m$ and $15~\mu m$ respectively. The impact parameter resolution for $3-5\ GeV$ momentum tracks is about $12~\mu m$. The angular acceptance of the vertex detector  averages to $\pm 250\ mrad$.

An original method of the charged particle track recognition and reconstruction in the spectrometer has been developed and successfully used at the first stage of the SVD experiment\cite{method}. This method was improved  for SVD-2 data analysis by using pre-calculated tables of possible particles trajectories in the magnetic field. The spectrometer with the vertex detector permits to obtain high effective mass resolution, for example, standart deviations in mass destributions are $4.4~MeV/c^2$ and $1.6~MeV/c^2$ for $K^0_s$-meson and $\Lambda^0$ masses respectively (see Fig.2 and Fig.3). The momentum resolution for the track with 15 measured hits is $(0.5\div1.0)\%$ for the $(4\div20)\ GeV/c$ momentum range. The angular measurement error is defined by the coordinate resolution of the vertex detector and by the effects of multiple coulomb scattering in the materials of the targets. This error was estimated to be $0.2\div0.3\ mrad$. The angular acceptance for the spectrometer averages to $\pm200\ mrad$ and $\pm150\ mrad$ for horizontal and vertical coordinates respectively.

The combined $(\pi^+K^0_s)$ and $(\pi^-K^0_s)$ invariant mass spectrum
is shown on Fig 4a. The $K^*(892)$ peak is clearly seen on the distribution. Fig 4b shows $(\Lambda^0 \pi^+)$ invariant mass spectrum. $\Sigma^+(1385)$ peak is clearly seen. The masses of observed  $K^0_s$, $\Lambda^0$ and also masses and widths of $K^*(892)$ and $\Sigma^+(1385)$ are consistent with their PDG values\cite{pdg}.

\begin{figure}[h]
\vspace{80mm}
{\includegraphics{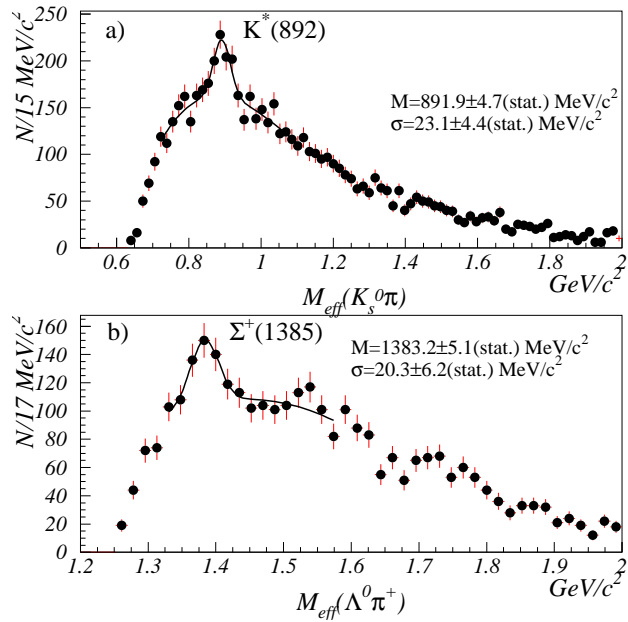}}
\caption{
a) The combined $(\pi^+K^0_s)$ and $(\pi^-K^0_s)$ invariant mass spectrum. b) $(\Lambda^0 \pi^+)$ invariant mass spectrum.}
\label{kstar}
\end{figure}

\section {$pK^0_s$-spectrum analysis.}

The events with multiplicity of no more than five charged  tracks (from primary vertex) in the vertex detector were selected to minimize the combinatorial background. This selection also reduces the probability of appearance of the events with rescattering on nuclei and also background of $K^0_s$-mesons produced in the central rapidity region. About $34\%$ of all inelastic events and $15\%$ of all detected $K^0_s$-mesons survived after using of this selection. It was estimated that for the selceted events the average multiplicity of produced particles taking in the account the neutral component ($\pi^0$-mesons and neutral strange particles) was about $8$ particles.

\begin{figure}[ht]
\vspace{80mm}
{\includegraphics{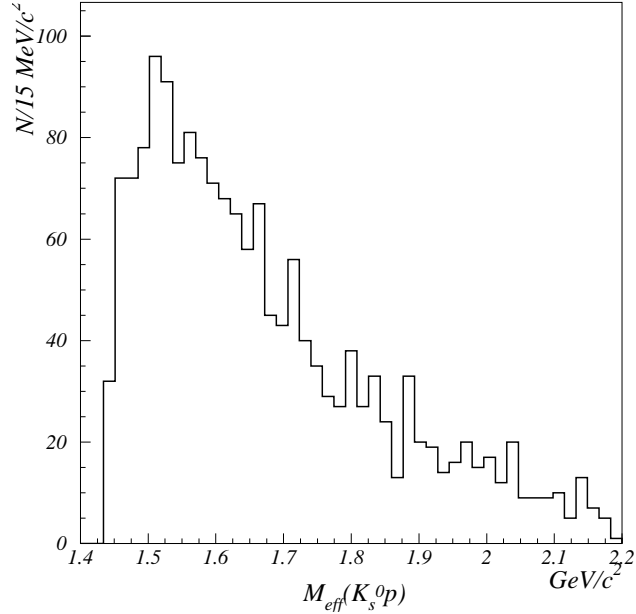}}
\caption{ The $(pK^0_s)$ invariant mass spectrum in the
reaction $pA\rightarrow pK^0_s+X$.}

\label{theta}
\end{figure}

$K^0_s$-mesons were identified by their charged decay mode $K^0_s \rightarrow \pi^+ \pi^- $, where oppositely charged tracks with high impact parameters were combined together. To eliminate contamination from $\Lambda^0$ decays, candidates with ($p \pi^-$) mass hypothesis less than $1.12\ GeV$ were rejected. The resulting invariant mass distribution is shown on Fig 2. About $3800$ $K^0_s$-mesons decayed within the vertex detector (decay length $\le 35\ mm$) were found in the selected events. The mean decay length of $K^0_s$-decays for all targets was found to be  $20\ mm$.

Protons were selected as positively charged tracks with number of spectrometer hits more than 15 and momentum $4\ GeV/c \le P_p \le 21\ GeV/c$. Pions of such energy should leave a hit in the Threshold Cherenkov counter, therefore absence of hits in TCC was also required.

Effective mass of the $pK^0_s$ system is shown on Fig 5. No obvious structure in this spectrum is seen, however there is a small enhancement in the $1530\ Mev/c^2$ mass region.

The following cuts have been applied:

\begin{itemize}
\item $490\ MeV/c^2 \le M_{\pi^+\pi^-} \le 505\ MeV/c^2$
\item $cos(\alpha) \ge 0$, where $\alpha$ is angle of flight of $pK^0_s$-system in the center mass system of the beam proton and the target nucleon.
\end{itemize}

The first of these cuts improves the resolution of the invariant mass of $pK^0_s$-system and the second one corresponds to the spectrometer aperture and reduces the background from misidentified protons.

Effective mass of the $pK^0_s$ system after these cuts is plotted in Fig. 6. A narrow enhancement is seen at the mass $M=1526\pm 3\ MeV/c^2$ with a $\sigma=10\pm 3\ MeV/c^2$. FRITIOF\cite{fritiof} simulations of minimum-bias $p-Si$ events at $70~GeV/c$ were
performed and analyzed using the same cuts in order to look for the
background events. Trigger and $K^0_s$ decay conditions, detector
acceptances and tracking uncertainties were also taken into account. It is seen that obtained FRITIOF background fails to reproduce the real background shape. This may be caused by the bunch of $\Sigma^{*+}$-bumps in the $(1560\div1800)~MeV/c^2$-mass area \cite{pdg}, which have rather high branching ratio for $(pK^0_s)$-decays. For $KN$-system some wide Deck-mechanism produced peaks may appear in this region of masses\cite{deck}. Therefore it was suggested to apply $P_{K^0_s}\le~P_p$ kinematical cut \cite{levtch}, because this cut efficiently destroys most of the decays of $\Sigma^{*+}$-resonances with high masses while cutting only $10\%$ of $\Theta^+$-events.

\begin{figure}[ht]
\vspace{80mm}
{\includegraphics{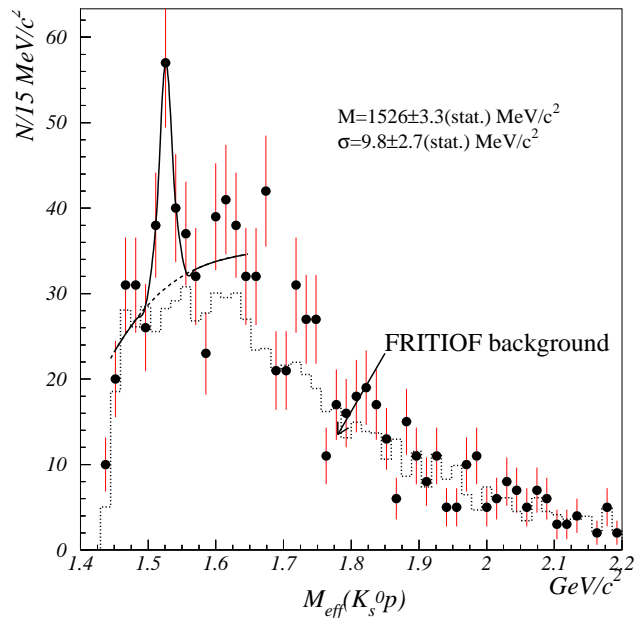}}
\caption{ The $(pK^0_s)$ invariant mass spectrum in the
reaction $pA\rightarrow pK^0_s+X$ with the cuts explained in text. Dashed histogram represents background obtained from FRITIOF simulations.}

\label{theta}
\end{figure}

\begin{figure}[ht]
\vspace{80mm}
{\includegraphics{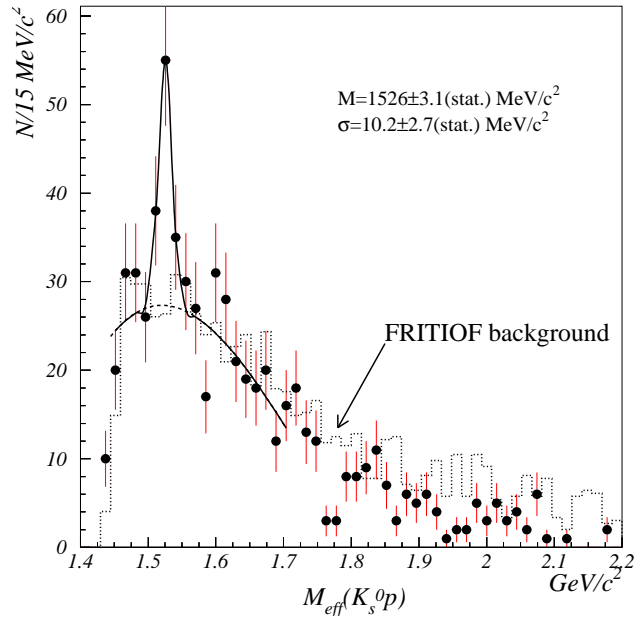}}
\caption{ The $(pK^0_s)$ invariant mass spectrum in the
reaction $pA\rightarrow pK^0_s+X$ with the additional kinematical cut explained in text. Dashed histogram represents background obtained from FRITIOF simulations.}

\label{theta}
\end{figure}

Resulting distribution is shown on fig. 7.
The distribution was fitted by Gaussian function and fourth-order polynomial background. The dashed histogram represents background obtained from FRITIOF simulations.  There are 50 events in the peak over 78 background events. The statistical significance for the fit on Fig 6. inside a $45\ MeV/c^2$ mass window is calculated as $N_{P}/\sqrt{N_{B}}$,
where $N_{B}$ is the number of counts in the background fit under the
peak
and  $N_{P}$ is the number of counts in the peak. We estimate the
significance
to be $5.6~\sigma$. It is impossible to determine the strangeness of this state in such inclusive reaction, however there are no reported $\Sigma^{*+}$-resonances in $1500\div1550~MeV/c^2$ mass area, so we interpret observed state as recently reported $\Theta^+$-baryon with positive strangeness.

It was verified that observed $pK^0_s$-resonance can not be a reflection from other (for example $K^{*\pm}(892)~or~\Delta^0$) resonances or artificially created peak.
No significant peaks were found in the $(pK^0_s)$ invariant mass spectra for events where $\pi^+$-meson was detected by TCC and its mass was substituted by proton mass. The $\pi^+$-background averages to no more than $10\%$ under selection criteria used.

Since the mechanism of the $\Theta^+$-production and dependence of the cross-section on charged particles multiplicity is unknown, one can only roughly estimate the full cross-section of the $\Theta^+$-production in the proton-nuclear interactions. The efficiency of detection of $\Theta^+$-baryon was estimated using simple Monte-Carlo model with variation of the $pK^0_s$-resonance energy spectrum. The knowledge of probability of $K^0_s$-decays within the vertex detector, the acceptance of the setup as well as probabilities of decays of $\Theta^+$ into $K^0_s$, $K^0_L$ and branching ratio of $K^0_s \rightarrow \pi^+\pi^-$ give the main contributions in this efficiency, and this efficiency was estimated to be $0.07\%$. The full cross section of $\Theta^+$-production in the proton-nuclear interactions is found to be $(30\div120)~\mu b$ (for $X_F\ge 0$ region) and this rather big dispersion is caused by uncertainty in cross section dependence on charged particles multiplicity, different number of observed events depending on different background models and detection efficiency.

The A-dependence analysis in the observed peak area showed that this dependence does not differ (within measuring errors) from the similar dependence for background inelastic events which is proportional to $A^{0.7}$. This result is contrary to the paper \cite{itep}, where
the strong A-dependence of $\Theta^+$-production in $\nu N$-interaction was claimed.

\section {Summary.}

In summary, the inclusive reaction $p A \rightarrow pK^0_s + X$ was studied at IHEP accelerator with proton energies at $70\ GeV$ using SVD-2 detector. With several cuts applied a narrow baryon resonance was observed with mass $M=1526\pm 3(stat.)\pm 3(syst.)\ MeV/c^2$ and $\Gamma < 24\ MeV/c^2$. The width of this state is close to SVD-2 experimental resolution for $pK^0_s$-system and its mass and width are consistent with recently reported $\Theta^+$-resonance, which was predicted as exotic pentaquark ($uudd\bar{s}$) baryon state\cite{Nakano, Dolgolenko,Ken,vpk,SAPHIR,itep,clas2,hermes}.

We wish to thank Dr. V. Kubarovsky (RPI/JLab), Dr. B. Levchenko (SINP MSU) and Dr. N. Zotov (SINP MSU) for useful comments and suggestions.

This work was supported by the following foundations: Russian Foundation for Basic Research (N 03.02.16894), The Program "Universities of Russia" (N 02.03.006/03-2), Russian Foundation for leading scientific schools (N 1685.2003.02) and contract with Russian Ministry of Industry, Science and Technology (Goskontrakt N 40.032.11.34) in the part of the development and creation of the vertex detector.

\end{document}